\documentclass{aa}  
\usepackage{newtxtext,newtxmath}
\usepackage[T1]{fontenc}
\usepackage{ae,aecompl}
\usepackage{natbib}
\usepackage{lipsum}
\usepackage{multirow}

\usepackage{graphicx}	
\usepackage{amsmath}	
\graphicspath{{./}{figures/}}
\usepackage{xcolor}
\usepackage{rotating}
\usepackage[normalem]{ulem}
\usepackage{lipsum}  
\usepackage{soul}
\usepackage{graphicx}
\usepackage{hyperref}
\hypersetup{
  colorlinks   = true, 
  urlcolor     = blue, 
  linkcolor    = blue, 
  citecolor   = blue 
}

\newcommand{\orcid}[1]{\textsuperscript{\href{http://orcid.org/#1}{
\hskip2pt\includegraphics[width=8pt]{Orcid-ID.png}}}}

\begin{document} 

\title{Completing the X-ray view of the recently discovered \\ supernova remnant G53.41+0.03}
\titlerunning{Completing the X-ray view of the recently discovered SNR G53.41+0.03}


   \author{V.~Domček\inst{1,2},
          J.~Vink\inst{1,2,3},
          P.~Zhou\inst{1,4},
          L. Sun\inst{1,2,4}
          \and
          L.~Driessen\inst{5}
          }

   \institute{Anton Pannekoek Institute for Astronomy, University of Amsterdam, Science Park 904, 1098 XH Amsterdam, The Netherlands\\
              \email{vdomcek@gmail.com}
    \and
    GRAPPA, University of Amsterdam, Science Park 904, 1098 XH Amsterdam, The Netherlands
    \and
    SRON, Netherlands Institute for Space Research, Utrecht, The Netherlands
    \and 
    School of Astronomy and Space Science, Nanjing University, Nanjing 210023, People’s Republic of China
    \and
    Jodrell Bank Centre for Astrophysics, School of Physics and Astronomy, The University of Manchester, Manchester, M13 9PL, United Kingdom
    }

   \date{Received xxx; accepted xxx}

 
  \abstract{
  \textit{Aims:} We present a detailed X-ray study of  the recently discovered supernova remnant (SNR) G53.41+0.03 that follows up and further expands on the previous, limited analysis of archival data covering a small portion of the SNR.  
  
  \textit{Methods:} With the new dedicated 70~ks {\it XMM-Newton} observation we investigate the morphological structure of the SNR in X-rays, search for a presence of a young neutron star and characterise the plasma conditions in the selected regions by means of spectral fitting.
  
  \textit{Results:} The first full view of SNR G53.41+0.03 shows an X-ray emission region well aligned with the reported half-shell radio morphology. We find two distinct regions of the remnant that differ in brightness and hardness of the spectra, and are both best characterised by a hot plasma model in a non-equilibrium ionisation state. Of the two regions, the brighter one contains the most mature plasma, with ionisation age $\tau \approx 4\times10^{10}$s~cm$^{-3}$ (where $\tau = n_e t$), a lower electron temperature of kT$_\mathrm{e} \approx 1$~keV and the highest estimated gas density of n$_\mathrm{H}\approx 0.87$~cm$^{-3}$. The second, fainter but spectrally harder, region reveals a younger plasma ($\tau \approx 1.7\times10^{10}$s~cm$^{-3}$) with higher temperature (kT$_\mathrm{e} \approx 2$~keV) and two to three times lower density (n$_\mathrm{H}\approx 0.34$~cm$^{-3}$). No clear evidence of X-ray emission was found for emission from a complete shell, the southern part appearing to be absent.
  Employing several methods for age estimation, we find the remnant to be $t \approx 1000-5000$~yrs old, confirming the earlier reports of a relatively young age. The environment of the remnant also contains a number of point sources, of which most are expected to be positioned in the foreground. Of the two point sources in the geometrical centre of the remnant one is consistent with the characteristics of a young neutron star. \\}
  




   {}


   \maketitle
%

\section{Introduction}

Supernova remnants (SNRs) are an important link in the cycle of gas in galaxies, connecting
star formation, stellar deaths and galactic chemical evolution. Moreover, the mechanical energy provided by supernovae (SNe) is transferred
to the interstellar gas by SNR shocks, which is an important ingredient for star formation \citep{Kim2015b,Koo2020}.
SNRs provide insights into the final stage of the lives of massive stars, as their shock waves illuminate the late mass loss history of
the SN progenitors, and their compositions, enhanced by freshly synthesised elements, reveal details about the explosion mechanism and progenitor properties.
Early on, after the explosion, they provide favourable conditions for dust condensation and their shock waves are thought to be the dominant producers of Galactic cosmic rays \citep{Vink2020Book}.
The interaction of the shocks and mixing of the SN produced elements with the interstellar medium lead to its chemical enhancement, thus providing SN feedback to the Galaxy. There is, therefore, a considerable interest in completing our view of the Galactic SNR population.

With the next generation of Galactic surveys in both radio---e.g. THOR \citep{Anderson2017}, the LOFAR Two-metre Sky Survey \citep{Shimwell2017}--- and X-rays ---e.g. eRosita \citep{Merloni2012}--- there has been a renewed interest in finding new SNRs. The reason is the still relatively low number of identified SNRs ($\sim$294 in \cite{Green2019} and $\sim$384 in \cite{Ferrand2012}) compared to the expected 2000--3000 estimated from the Galactic SN rate \citep{Vink2020Book}. 

G53.41+0.03 is one of the recently found SNRs that came out of mapping of the relatively crowded field along the Sagittarius-Carina and Perseus arms of the Galaxy. While it appeared as a candidate SNR in the THOR survey \citep{Anderson2017}, the measured radio spectral index of $\alpha \approx -0.6$ (for $S_v \propto \nu^\alpha$) from a LOFAR wide field investigation \citep{Driessen2018} confirmed its SNR nature. A portion of the remnant was covered by an archival {\it XMM-Newton} observation at the edge of one of its MOS detectors. An analysis of this partial data-set proved the existence of a hot non-equilibrium ionisation (NEI) plasma coinciding with the radio emission, further confirming the SNR nature of G53.41+0.03 and hinting at a relatively young age of the remnant for the estimated distance of about 7.5~kpc \citep{Driessen2018}.

In this work, we follow up SNR G53.41+0.03 with a new dedicated 70~ks {\it XMM-Newton} observation. We provide a first complete look on the morphology in X-rays and analyse three distinct regions of the remnant. We characterise the plasma conditions in these regions with an aim to provide a better understanding of the local plasma characteristics and more precisely estimate the age of the remnant. We also discuss potential neutron star candidates seen in the SNR's geometrical centre. 

\section{Observations and data reduction}
We performed a dedicated 70~ks {\it XMM-Newton} observation of the SNR G53.41+0.03 on 13th of October 2019 (ObsID: 0841190101, PI:~Domček). We used the new dataset together with the archival one (ObsID: 0503740101, PI:~Wang), which encompasses only part of the remnant at the edge of field of view (FoV) of one of the detectors \citep{Driessen2018}. 

\begin{figure*}
    \centering
    \includegraphics[width=0.98\columnwidth]{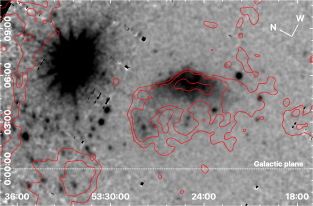}
    \includegraphics[width=0.98\columnwidth]{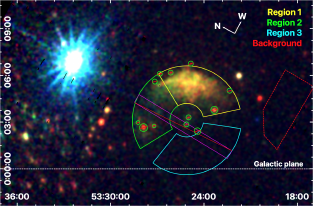} 
    \caption{SNR G53.41+0.03 with a type I outburst of HMXB IGR J19294+1816 in October 2019. Left: Grey-scale image in the energy range of 0.8-4.0~keV. Radio contours of the THOR VGPS survey \citep{Anderson2017} with levels -- (0.013, 0.01475, 0.02) Jy/beam are overlaid on top of the X-ray data. Right: Three-colour image (red 0.8--1.5~keV, green 1.5--2.5~keV and blue 2.5--4.0~keV) with extraction and exclusion regions used in the analysis. We applied square-root scaling on the left image and logarithmic scaling on the right image to improve visibility.}
    \label{fig_g53b:SNR_regions}
\end{figure*}

We extracted images and spectra with the Science Analysis System (SAS) v18.0. The tasks {\it epchain/emchain} were used for the data reduction while {\it pn-filter/mos-filter} were employed for filtering the background flaring. This resulted in clean exposure time of 52ks/65ks for MOS1/MOS2 and 42ks for pn in the case of 2019 observation and 32ks for MOS2 in the case of 2008 observation. Tasks {\it eimageget} and {\it eimagecombine} were used in order to obtain the image of the remnant in Fig.~\ref{fig_g53b:SNR_regions}. We used XSPEC (12.11.1) \citep{Arnaud1996} for our spectral analysis and modelling. We employed C-statistics \citep{Cash1979} and the optimal binning method {\it ftgrouppha} \citep{Kaastra2016}, which is part of the Heasoft package \footnote{\url{https://heasarc.gsfc.nasa.gov/lheasoft/}}.

\begin{table}
    \centering
    \begin{tabular}{ccc}
    ObsID & Date & Exposure time  \\
    \hline
    \hline
    0503740101 & 2008-03-29 & 58~ks \\
    0841190101 & 2019-10-13 & 70~ks \\
    \end{tabular}
    \caption{List of {\it XMM-Newton} observations used in this work.}
    \label{tab_g53b:observation_list}
\end{table}

Throughout the 2019 observation, a type I outburst from a nearby high mass X-ray binary was detected in the same FoV \citep{Domcek2019ATel}. This has caused an unexpected contamination of the SNR region. The problems are mainly present in the pn detector, although the MOS chips required the exclusion of affected regions as well. For that reason we focused our analysis on using the higher spectral resolution MOS spectra.

\section{Results}

\subsection{Morphology of the remnant}
\label{sec_g53b:results_morphology}
SNR G53.41+0.03 has a half-shell morphology of $3.5'$ in size with most of the emission coming from the upper half (in Galactic coordinates), as shown in Fig.~\ref{fig_g53b:SNR_regions}. 
The lower half of the SNR does not show any clear morphological detection in Fig.~\ref{fig_g53b:SNR_regions}. 
The SNR is located in an environment with several point sources in the line of sight. Two of these are particularly interesting, as they are positioned in the assumed geometric centre of the remnant. We discuss their properties and relation to the remnant further in section~\ref{sec_g53b:discussion_point_sources}. 

\subsection{Spectral analysis}
For our spectroscopic analysis, we divided the SNR 
into three regions based on the morphology and hardness of the spectra (see Fig.~\ref{fig_g53b:SNR_regions}, right). Region 1 was selected in the top part of the SNR, where the X-ray emission is brightest and its spectrum also appears softest. This region spatially coincides with the region previously analysed by \citep{Driessen2018}. Region 2 is positioned in the previously unobserved part of the remnant. This section is fainter and has a harder spectrum compared to region 1. Region 3 does not show obvious emission in the X-ray map but fills the spherical structure of the remnant. All three regions, including the background modelling region, are shown in Fig.~\ref{fig_g53b:SNR_regions} (right). 

We fit all background spectra in the wider 0.3-10.~keV range, while the 0.5-10.~keV energy range is used for the spectra of source regions. Uncertainties are reported in the 1$\sigma$ confidence range and are calculated through the Markov Chain Monte Carlo (MCMC) method, utilising the Goodman-Weare algorithm \citep{Goodman2010} with 600 walkers and 10$^6$ steps. The results of the MCMC runs are also provided in the Figures \ref{fig_g53b:mcmc_region1}, \ref{fig_g53b:mcmc_region2}, \ref{fig_g53b:mcmc_region2_2},  \ref{fig_g53b:mcmc_ps1} in the appendix of the paper.

\begin{table*}
    \centering
    \begin{tabular}{llc|ccc}
    \cline{3-5}
     &  & \multicolumn{3}{c}{Region/Model}  \\
    \cline{3-5}
     &  & 1 & \multicolumn{2}{c}{2} \\
    \hline
    Parameter & \multicolumn{1}{l|}{Units} & \textit{TBabs*vnei}  & \textit{TBabs*vnei} & \textit{TBabs*vnei} (5keV cut) \\
    \hline
    N$_\mathrm{H}$ & \multicolumn{1}{l|}{10$^{22}$ cm$^{-2}$} & 2.87$^{+0.11}_{-0.05}$   
    & 2.8$^{+0.4}_{-0.2}$ & 2.8$^{+0.3}_{-0.2}$  \\  
    kT$_\mathrm{e}$ & \multicolumn{1}{l|}{keV} &  0.97$^{+0.03}_{-0.03}$  
    & 2.9$^{+0.2}_{-0.2}$ & 1.9$^{+0.2}_{-0.2}$ \\    
    $\tau$ & \multicolumn{1}{l|}{$ 10^{10}$ s~cm$^{-3}$} &  4.2$^{+0.5}_{-0.4}$   
    & 1.5$^{+0.2}_{-0.2}$ & 1.7$^{+0.4}_{-0.3}$  \\   
    Ne & \multicolumn{1}{l|}{} &  1.0$^{+0.3}_{-0.1}$    
    & =Mg & =Mg   \\  
    Mg & \multicolumn{1}{l|}{} &  0.94$^{+0.13}_{-0.06}$  
    & 1.2$^{+0.5}_{-0.2}$ & 0.8$^{+0.5}_{-0.1}$ \\    
    Si & \multicolumn{1}{l|}{} &  0.54$^{+0.05}_{-0.20}$   
    & 0.88$^{+0.2}_{-0.1}$  & 0.68$^{+0.18}_{-0.05}$  \\   
    S & \multicolumn{1}{l|}{} &  1.06$^{+0.11}_{-0.07}$   
    & 0.8$^{+0.3}_{-0.2}$ & 0.7$^{+0.3}_{-0.2}$ \\ 
    Fe & \multicolumn{1}{l|}{} &  0.9$^{+0.2}_{-0.1}$  
    & 1.0$^{+1.1}_{-0.4}$ & 0.7$^{+1.0}_{-0.2}$  \\   
    norm & \multicolumn{1}{l|}{10$^{-4}$} &  55$^{+4}_{-4}$  
    & 4.6$^{+1.0}_{-0.4}$ & 7$^{+1}_{-1}$  \\   
    \hline
    F$_\mathrm{(1-10)keV}$ & \multicolumn{1}{l|}{$ 10^{-12}$ erg~cm$^{-2}$s$^{-1}$} & 5.9$^{+0.6}_{-0.5}$ 
    & 1.2$^{+0.3}_{-0.3}$ & 1.1$^{+0.5}_{-0.3}$ \\
    \hline
    \multicolumn{2}{c|}{Cstat/(d.o.f.)} & 329/277 & 195/169 & 86/104  \\
    \end{tabular}
    \caption{Best-fit values for 2 selected regions of the remnant. 1$\sigma$ confidence range was calculated using XSPEC's Goodman-Weare MCMC method with 10$^6$~steps and~600 walkers. 1$\sigma$ uncertainties for the unabsorbed flux (F$_\mathrm{(1-10)keV}$) were calculated using \textit{cflux} model and non-MCMC error calculation method provided by {\sc XSPEC}. \textbf{Bold} typeset signifies a fixed parameter.}
    \label{tab_g53b:models}
\end{table*}

\begin{figure*}
    \centering
    \includegraphics[width=\columnwidth]{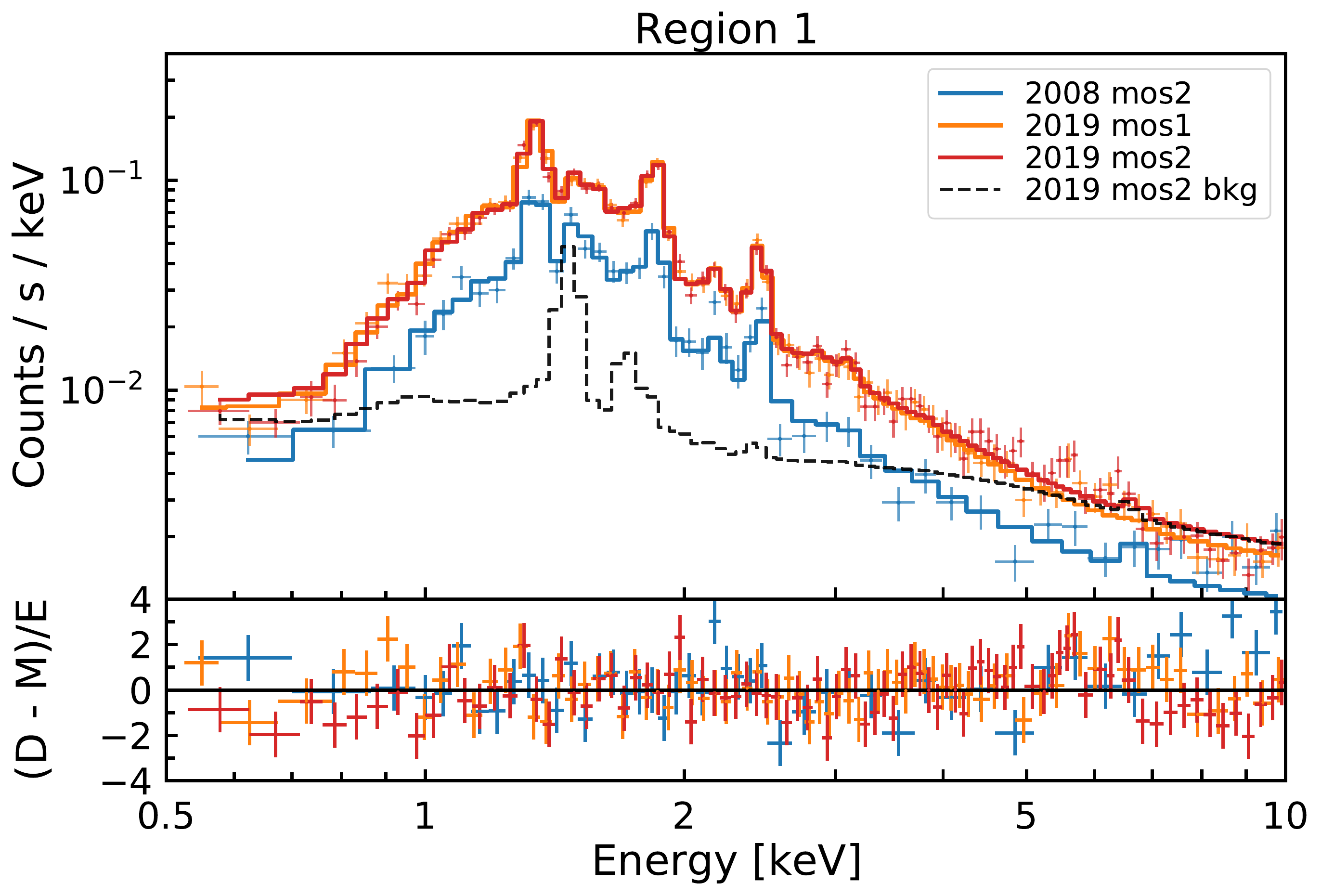}
    \includegraphics[width=\columnwidth]{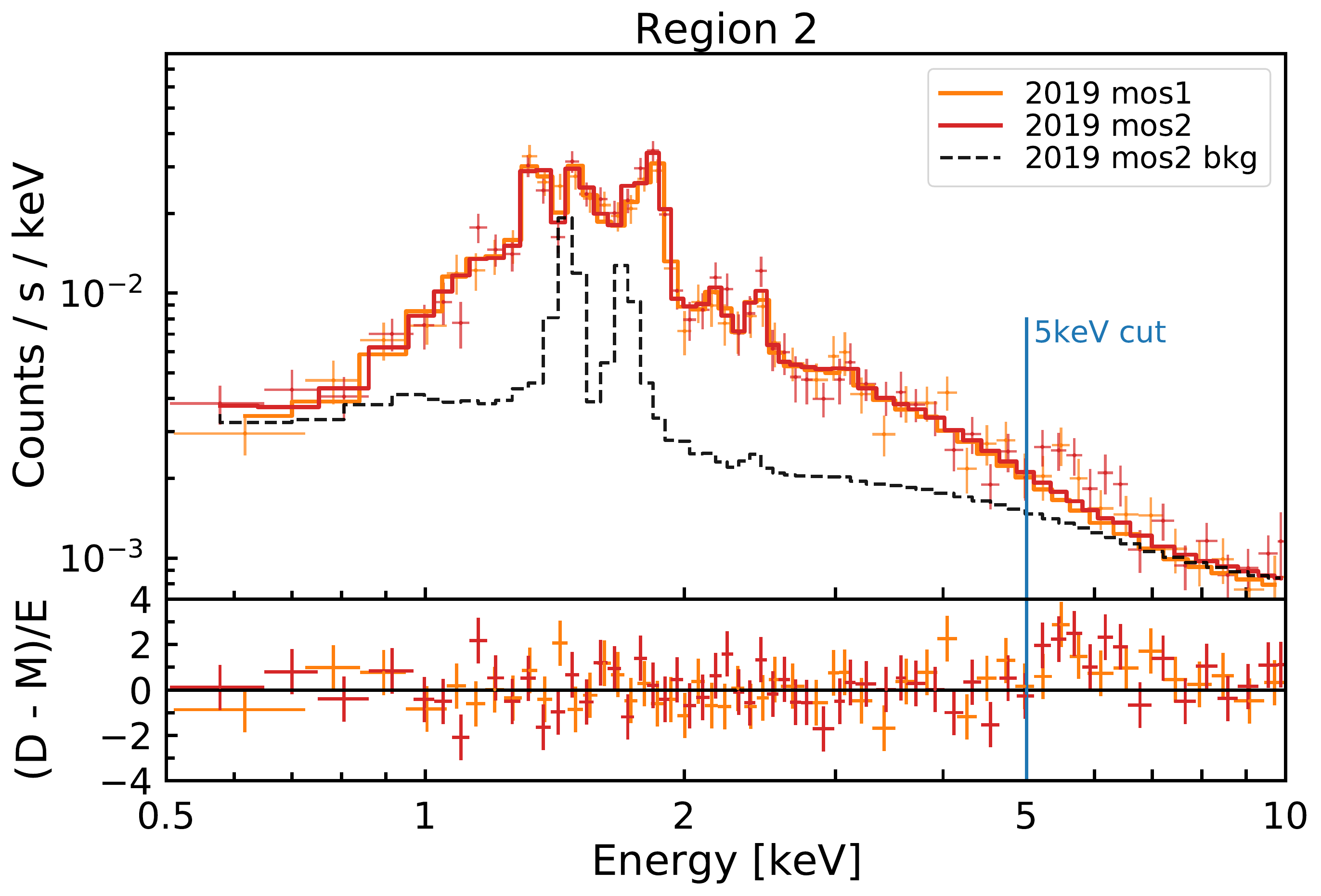}
    \includegraphics[width=\columnwidth]{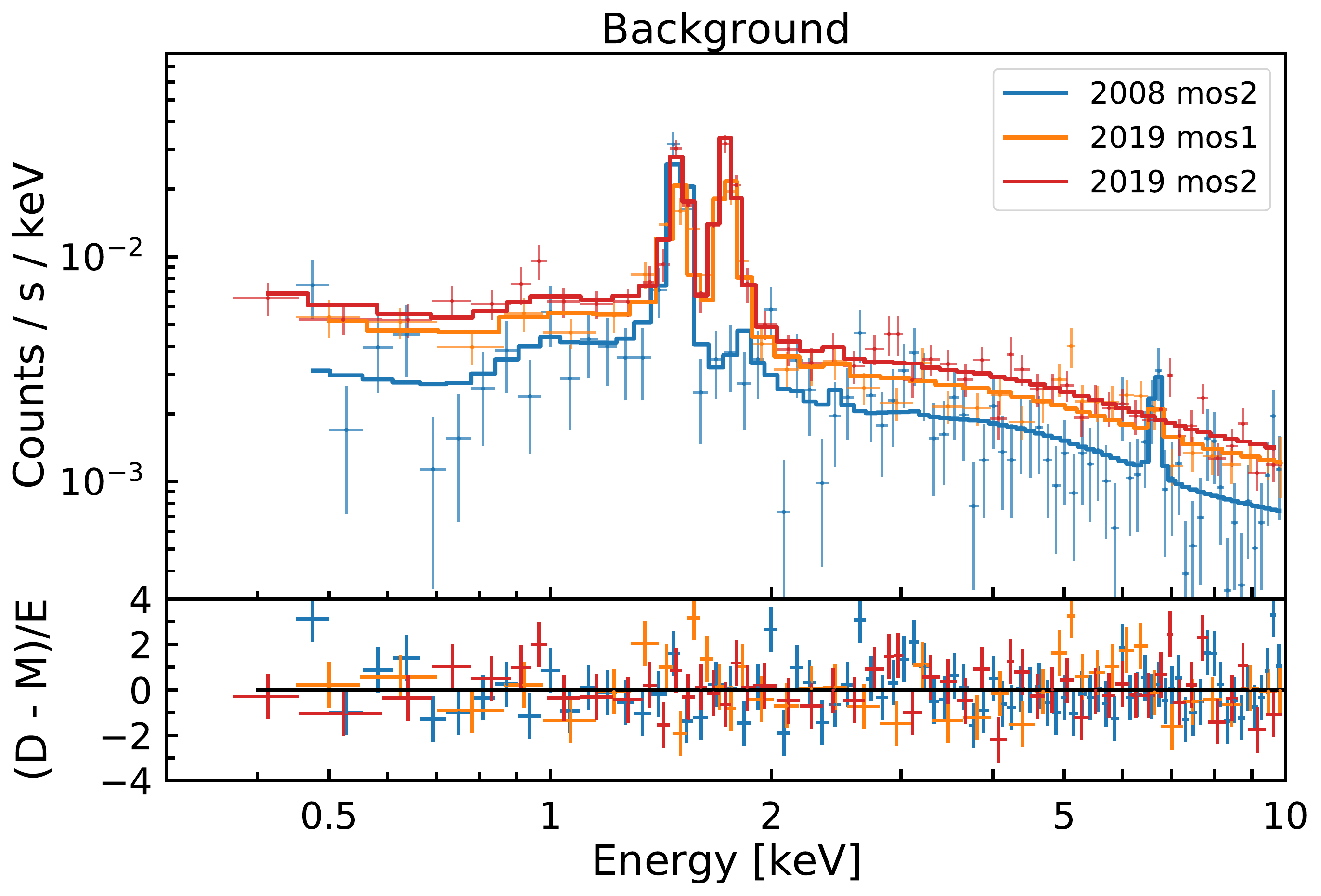}
    \caption{Best fit models for the source and the background region spectra. A black dashed line shows the MOS~2 detector background model for individual regions. In region 2 we display the best-fit model fitted in the 0.5-5.0~keV energy range.}
    \label{fig_g53b:spectra}
\end{figure*}

\subsubsection{Background modelling}
We modelled the extracted MOS1 and MOS2 backgrounds using a combination of the models {\it TBabs}, {\it Apec}, and {\it Power-law\footnote{
\href{https://heasarc.gsfc.nasa.gov/docs/xanadu/xspec/manual/XSmodelTbabs.html}{\textit{TBabs}} is the Tuebingen-Boulder ISM absorption model used for foreground Galactic absorption, 
\href{https://heasarc.gsfc.nasa.gov/xanadu/xspec/manual/XSmodelApec.html}{\textit{Apec}} is a thermal plasma model used for galactic diffuse emission, and 
\href{https://heasarc.gsfc.nasa.gov/xanadu/xspec/manual/XSmodelPowerlaw.html}{\textit{Power-law}} represents cosmic X-ray and particle backgrounds.
} } and several Gaussian lines. The Gaussian lines represent the instrumental background lines, and their normalisation is left to vary even for the source spectra due to their non-flat distribution across the detector. More details on the background model components are available in \cite{Masui2009} and \cite{Sun2020}.

The parameters of these model components are, apart from normalisation of the Gaussians, linked together for all of the MOS spectra in use. We obtained an acceptable fit with Cstat/d.o.f of 306/247. The spectrum together with the model are shown in the top-left panel of Fig.~\ref{fig_g53b:spectra}. 
This background model, scaled to the extraction area, was applied to all the subsequent MOS spectra, except for the central point source analysis that has a separate background model treatment.

\subsubsection{Region 1}
The spectrum is well represented with a single non-equilibrium ionisation plasma model {\it vnei}\footnote{\url{https://heasarc.gsfc.nasa.gov/xanadu/xspec/manual/XSmodelNei.html}} \citep{Borkowski2001} with the abundances of Ne, Mg, Si, S and Fe released. The best-fit values with 329/277 (Cstat/d.o.f.) show the plasma temperature to be kT$_\mathrm{e}= 0.97 ^{+0.03}_{-0.03}$~keV, the ionisation age $\tau = 4.2 ^{+0.3}_{-0.4} \times 10^{10}$~s~cm$^{-3}$ and mostly solar abundances being consistent with the previous findings for this particular region. The total number of counts, including background, for this region is $\sim$19200 ct.

\subsubsection{Region 2}
Region 2 is positioned in the previously uncovered northern part of the remnant and, therefore, only MOS1/2 spectra from the 2019 observation are available. 
Applying the non-equilibrium model {\it TBabs*vnei} reproduces the spectrum well with the fit statistics of 197/171 (Cstat/d.o.f.). Although the plasma temperature is expected to be higher based on the hardness of the three-colour image, the fitted temperate kT$_\mathrm{e}= 2.9 ^{+0.5}_{-0.4}$~keV is still rather high. 

We suspect that the higher temperature compared to region 1 could be a flux excess in the spectrum around 5--6~keV (see lower-left panel of the Fig.\ref{fig_g53b:spectra}). We considered two explanations for this excess: i) a physical interpretation, which would explain the excess with non-thermal emission coming out of this part of the SNR; or ii) a technical interpretation, where the reason for the excess could be explained with a local calibration or background issue.

For the first scenario, we included an additional power-law component with $\Gamma = 3$ into our model. 
We obtained a marginally better fit to the data with 191/170 (Cstat/d.o.f.). However, the plasma temperature 1$\sigma$ confidence range increased to an even higher, and less probable, kT$_\mathrm{e}= 5 ^{+7}_{-3}$~keV. 
Additionally, releasing the power-law index leads to a preferred value of $\Gamma \approx 0.3$, a value that is not consistent with the range of values known for the non-thermal emission in SNRs, as reported for example in \cite{Sasaki2004}, Table.~3.

The alternative explanation is contamination from quiescent particle background (QPB), where Cr and Mn lines are known to be present in this energy range \citep[][see Fig.~2 and Fig.~3]{Kuntz2008}. 
We therefore restricted the fit to data only in the energy range of 0.5-5.0~keV, where the SNR's emission is clearly dominating. The fit statistics in this case is 86/104 (Cstat/d.o.f.), with the fitted temperate moving to a lower estimate of kT$_\mathrm{e}= 1.9 ^{+0.2}_{-0.2}$~keV. The best-fit ionisation age is $\tau = 1.7 ^{+0.4}_{-0.3} \times 10^{10}$~s~cm$^{-3}$ and abundances stay roughly solar and similar for both energy ranges. Full details on the fitted models can be found in Table~\ref{tab_g53b:models}. The total number of counts, including background, for this region is $\sim$4200 ct.

\subsubsection{Region 3}
\label{sec_g53b:results_reg3}
Spectrum of region 3 does show a minor surplus of emission above the background model, mainly in the 1.2--1.4~keV energy range. However, considering the faintness of the emission and the variability in the background across the field of view, we do not find sufficient evidence for significant detection. The total number of counts, including background, for this region is $\sim$5100 ct.

\section{Discussion}
Previous work by \cite{Driessen2018} confirmed the presence of an X-ray emitting plasma coinciding with the location of a radio structure that could be explained only by SNR shocks. The probable distance of the remnant was placed to the outer edge of the Sagittarius-Carina arm at approximately 7.5~kpc. Using a various age estimation techniques, the age of the SNR was estimated to be between 1000 and 8000 years. With the newly acquired data-set we build upon the previous results and provide a more complete picture of the remnant.

\subsection{Visible structure of the SNR}
A full view of G53.41+0.03 shows a half-spherical X-ray morphology with the size of the remnant somewhat smaller ($\theta_X \sim3.5'$) than the size estimated in the radio band ($\theta_R \sim 5'$). However, the difference in size estimate is most likely caused by a better image resolution in the X-ray band and not by a different morphology. We can therefore now confirm that the overall X-ray morphology aligns well with the visible radio structure. 

The more accurate measurements of the SNR's angular size result in updates of the previously reported characteristics \citep{Driessen2018}, such as estimates of physical size, number densities and the plasma age. For the distance estimate of $d \approx 7.5$~kpc \citep{Driessen2018}, the physical radius of the SNR is $r \sim 7.64 d_{7.5}$~pc, where $d_{7.5} = d/7.5$~kpc. 

As mentioned in section \ref{sec_g53b:results_morphology}, the {two} regions selected for {further} analysis differ in the brightness and spectral characteristics. These are likely caused by density variations in the SNR and its ambient medium.
We test this idea by estimating the number densities in those regions.

In order to calculate the number densities for the individual regions ($n_{\rm H,i}$), we use 
\begin{equation}
    n_\mathrm{H,i} = \sqrt{\frac{4 \pi 10^{14} \left( \frac{d_{7.5}}{\mathrm{cm}} \right)^2 \times \mathrm{norm}_i}{1.2 \ V_{i}}},
    \label{eq_g53b:density}
\end{equation}
where norm$_i = (10^{-14}/4\pi D^2) \int n_\mathrm{e} n_\mathrm{H} dV$ is a normalisation per region\footnote{\cite{Driessen2018} had an incorrect n$_\mathrm{H} \propto d_{7.5}^{-3/2}$ dependence due to a missed normalisation factor.}, $d_{7.5}$ is the distance to the SNR and $V_i$ is the volume of emitting plasma. We take $n_{\rm e}=1.2n_{\rm H}$ to account for the electron number densities $n_{\rm e}$ released by the ionisation of hydrogen atoms.

To estimate the total volume of the emitting plasma we assume a Sedov remnant with a shock compression ratio of $\chi \sim 4$. The volume-filling factor is in that case $\sim 25\%$ \citep[e.g.][]{Reynolds2017}, leading to estimate of $V_\mathrm{SNR} = 1.37 \times 10^{58}d^3_{7.5}$cm$^3$. The total volume is distributed in individual regions according to their size (Table~\ref{tab_g53b:age_calculations}).

\begin{table}
    \centering
    \begin{tabular}{ll|cc}
        \hline
        Parameter & Units & Region 1 & Region 2 \\
        \hline
        Volume & [$\%$] & 30 & 25  \\
        Volume $\times f_{0.25}$ & [cm$^3$] & 4.11$\times10^{57}$ & 3.42$\times10^{57}$ \\
        n$_\mathrm{H} \times f_{0.25}^{-1/2} \times d_{7.5} ^{-1/2}$ & [cm$^{-3}$] & 0.87 & 0.34 \\
        kT$_\mathrm{e}$ & [keV] & 0.97 & 1.87  \\
        v$_{sh}\times \sqrt{\beta}$ & [km~s$^{-1}$] & 900 & 1247  \\
        \hline
        \multicolumn{4}{c}{Age estimates}\\
        \hline
        $t_v\times d_{7.5} \times \sqrt{\beta}$  & [kyr] & 3.3  & 2.4 \\
        $t_E\times d_{7.5} ^{9/4} \times f_{0.25}^{-1/4}$   & [kyr] & 1.4  & 0.8 \\
        $t_{\tau _0} \times d_{7.5} ^{1/2} \times f_{0.25}^{1/2} \times k_{0.3} $  & [kyr] & 5.2 & 5.1  \\
    \end{tabular}
    \caption{Volume of emitting plasma, number density and plasma age estimates for regions in SNR G53.41+0.03. Parameters $t_v$, $t_E$ and $t_{\tau _0}$ represent plasma age estimates calculated using the Sedov shock velocity, Sedov radius and ionisation age methods respectively. 
    $ d_{7.5} = (d/7.5\mathrm{kpc})$ is the distance of 7.5~kpc, $\beta$ is the ratio between post-shock electron and proton temperature (see sec.~\ref{sec_g53b:sedov-taylor}), $f_{0.25} = (f/0.25)$ is the volume-filling factor normalised to 25\%} while $k_{0.3} = (k/0.3)$ represents correction for the difference between the measured bulk of the plasma ionisation age ($\tau$) and the maximum possible $\tau _0$ present in the plasma (see sec.~\ref{sec_g53b:ionisation_age}). 
    \label{tab_g53b:age_calculations}
\end{table}

The calculated particle densities show, not surprisingly, 
that the brighter region 1 also has higher density.
Region 1 consists of $2-3 \times$ denser material compared to region 2 (Table~\ref{tab_g53b:age_calculations}). This leads to faster particle interaction and cooling, hence stronger X-ray emission. In this scenario we could also expect the plasma in region 1 to be more mature, with higher values of $\tau$, and lower temperature kT$_\mathrm{e}$ due to the plasma cooling efficiency. This is indeed what we see in Table~\ref{tab_g53b:models}.

A possible reason for the {difference in densities} could be that it arises from an asymmetry in the stellar wind of the progenitor star, caused by the
movement of the star with respect to the local interstellar medium.
The idea was originally proposed by \cite{Weaver1977} and has been evoked, for example, in the case of RCW 86 \citep{Vink1997}. More recent 2D simulations of wind bubbles produced by runaway stars have been calculated among others by \cite{Meyer2015}. The authors show an explosion of a supernova in an asymmetric bubble environment and predict that while the SNR explosion wave reaches accumulated matter of the bow shock within the first $\sim 1000$~yr, the collision with the rest of the cavity border continues for several thousand years afterwards. Although specific progenitor star parameters might differ, we could nevertheless be observing the second stage of this scenario, in which the shock wave has already reached the bow shock over-density and is just now expanding through the lower density cavity border. Another alternative could be an interstellar medium (ISM) enhancement unrelated to the evolution of the progenitor star.

The stellar wind asymmetry model or the ISM enhancement could explain why {region 2 and 3} show much fainter or no emission compared to region 1, though it might not be the only reason for the visual distinction. 
For example, region 3 also lies much closer to the galactic plane and thus experiences much higher line of sight absorption, as indicated by the higher measured hydrogen column density N$_\mathrm{H}$. This is supported by the presence of the infrared dark cloud (IRDC) G53.2 with associated CO emission at distance of 1.7~kpc \citep[Fig.~3 and 9]{Kim2015a}.

\subsection{Age of the remnant}
There are several approaches to estimating the age of X-ray emitting plasma in SNRs. The most widely used ones are motivated by either the Sedov-Taylor self-similar evolution model 
or based on the degree of ionisation non-equilibration of elements with prominent emission lines. We will discuss both methods here.

\subsubsection{Sedov-Taylor model}
\label{sec_g53b:sedov-taylor}
In the Sedov-Taylor approach, the plasma age can be estimated from the fitted parameter of the post-shock electron temperature kT$_\mathrm{e}$. 
Under the assumption of full equilibrium of electron and ion temperatures, kT$_\mathrm{e}$ ties to the shock velocity $v_{sh}$ as
\begin{equation}
    \mathrm{kT}_e \sim 1.2\left( \frac{v_{sh}}{1000 \mathrm{\ km \ s^{-1}}} \right)^2 \mathrm{keV}.
    \label{eq_g53b:temperature_equilib}
\end{equation}
The obtained shock velocity is further related to the plasma age $t_{v}$ through the Sedov-Taylor model as
\begin{equation}
    t_{v} = 0.4r/v_{sh},
\end{equation}
where $r$ is the size of the remnant.
Note that the measured $T_e$ and the post-shock proton temperature $T_p$, which dominates the internal energy of the plasma, are not necessary the same. \cite{Vink2015} showed how the ratio $\beta = T_e/T_p$ depends on the shock's Mach number and in case of the SNRs can range from $\beta = 0.05$ to $1.0$. Lower $\beta$ values can result in lower age estimates by a factor $\sqrt{\beta}$. For comparison, \cite{Sasaki2004} used $\beta = 0.4$ in the case of more mature CTB 109. If we used $\beta = 0.3$ for a presumably younger G53.41+0.03, $t_v$ estimates in Table.~\ref{tab_g53b:age_calculations} would be around half of those calculated for the full equilibrium.

Another independent age estimate based on the Sedov-Taylor model depends on the SNR's radius $R$, pre-shock density $\rho =(n_\mathrm{H}/\chi) \times m_p$, where $\chi$ is a shock compression ratio, and $E$ the explosion energy 
\begin{equation}
    t_{E} = \sqrt{\frac{1.4(n_\mathrm{H}/\chi) m_p R^5}{2.026 E}}.
\end{equation}
We use the updated radius $R=7.64$~pc and canonical explosion energy $E=10^{51}$~erg. The post-shock number densities $n_\mathrm{H}$ (Table~\ref{tab_g53b:age_calculations}) are converted into pre-shock densities required for this calculation by assuming a strong shock compression ratio of $\chi=4$. The resulting pre-shock densities are used as lower and upper bounds for the age estimation. We find the age to be in the range of $t_E \sim 850 - 1350$~yrs. It is likely that the average estimate lies somewhere in between these two values. 

\subsubsection{Ionisation age}
\label{sec_g53b:ionisation_age}
Ionisation age $\tau$ is a measure of ionisation non-equilibrium in a plasma. It is a product of the plasma (number) density $n_e$ and the time since the material has been shocked $t_\tau$. Given the estimates of $n_\mathrm{H}$ provided in Table.~\ref{tab_g53b:age_calculations}, and using our previous assumption of $n_e = 1.2 n_\mathrm{H}$, we can obtain another independent estimate of the plasma age based on $t_{\tau}= {\tau}/{n_e}$. 

However, the SNR shell contains plasma shocked fairly recently at $t\approx 0$, as well as plasma that has been shocked not long after the supernova explosion.
Moreover, during the expansion the density changes, so for a given plasma element what counts is $\tau=\int n_{\rm e}(t)dt$. 
The measured $\tau$ characterises the bulk of the plasma ionisation age and is necessarily smaller than the maximum possible $\tau_0 \approx n_{\rm e}t_{\tau _0}$,
with $t_{\tau _0}$ being the actual age of the SNR.
\cite{Borkowski2001} reported that in the case of the Sedov-Taylor model with the uniform ambient ISM distribution, the most common $\tau$ within the shell has $\tau \approx 0.3 {\tau _0}$.

The conversion of $\tau$ to age is dependent on the density distribution around the SNR, which is likely affected by the mass loss history of the SNR progenitor. For example, a stable stellar wind has $\rho \propto r^{-2}$, leading to a higher density near the explosion centre, and hence a larger $\tau$, if we compare the age and the density near the shock radius. 

\begin{figure*}
    \centering
    \includegraphics[width=\columnwidth]{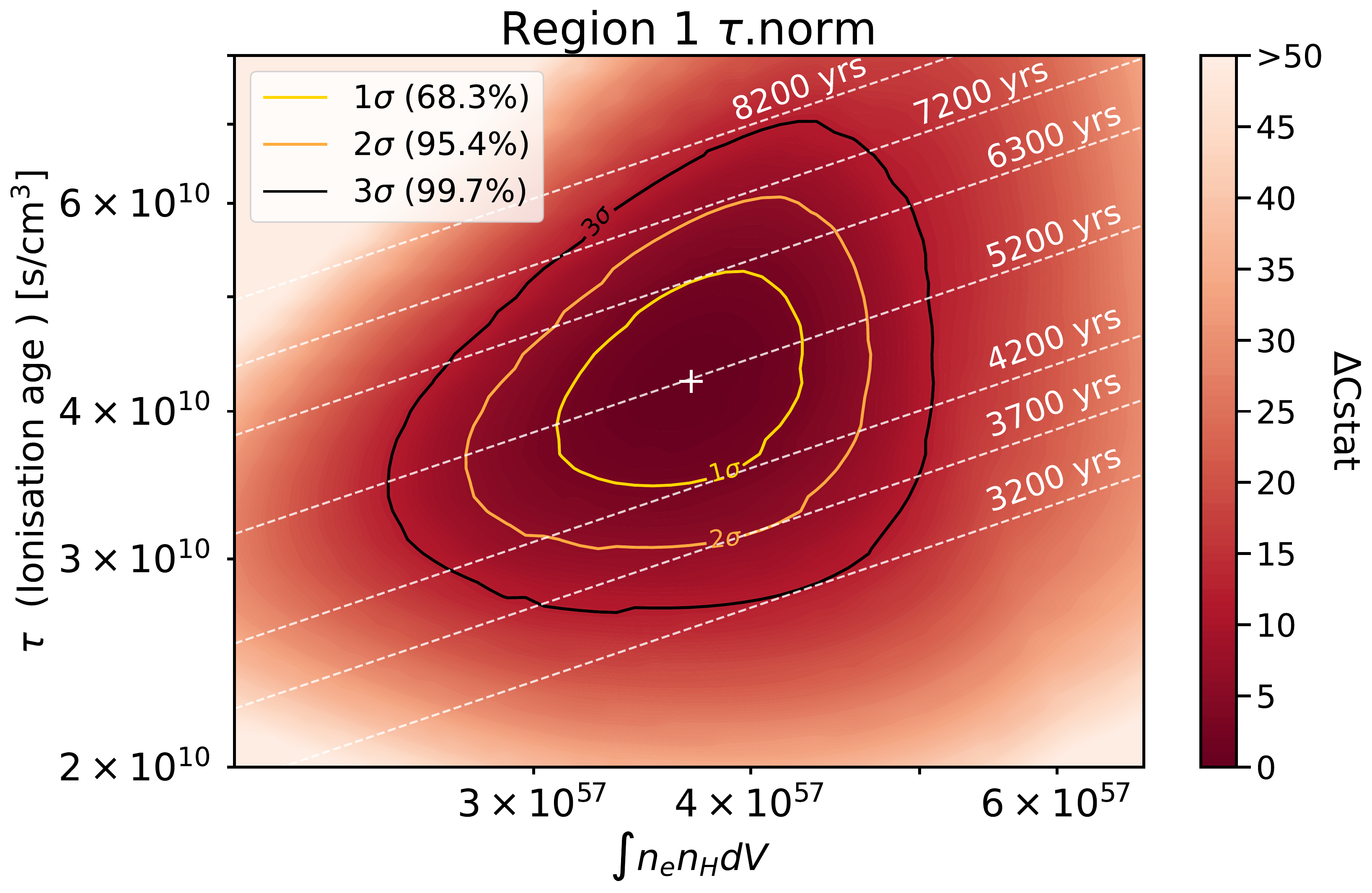}
    \includegraphics[width=\columnwidth]{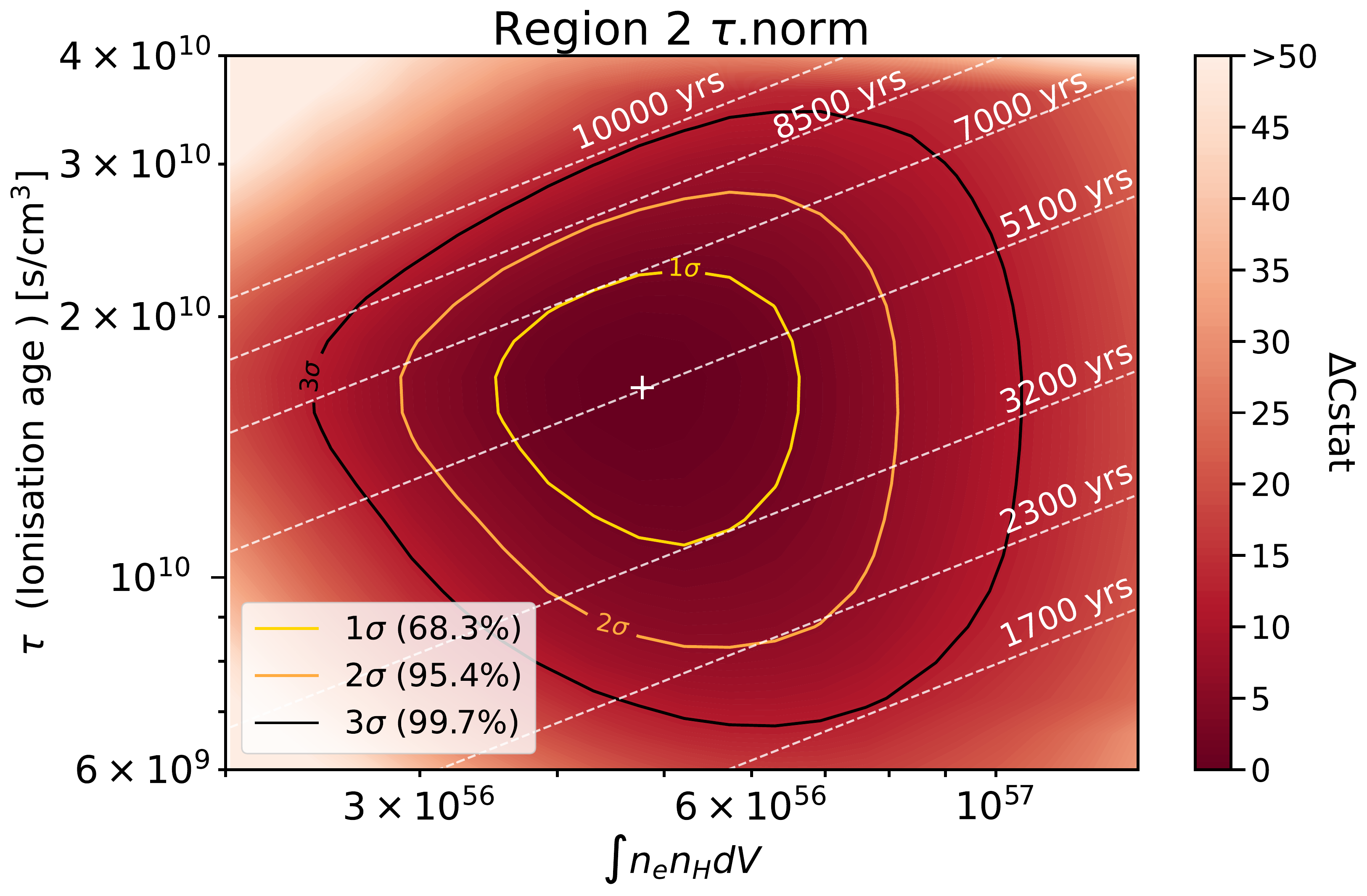}
    \caption{2D contour plots of emission measure vs ionisation age $\tau$. Dashed lines represent curves of the same ($\tau/n_{\rm e}$) prescribed by Eq.~\ref{eq_g53b:time_curves}.}
    \label{fig_g53b:cont_tau_n}
\end{figure*}

The two regions (1 and 2) show a consistent picture with age of the shocked plasma being around $\approx 5.2d^{1/2}_{7.5}$ kyrs old. 
To understand the range of allowed ages 
we produced contour plots between the emission measure ($\int{n_\mathrm{e} n_\mathrm{H} dV} = 4\pi \times 10^{14} \times \left({d_{7.5}}/{\mathrm{cm}}\right)^2 \times $norm$_i$) and ionisation age ($\tau$) in Fig.~\ref{fig_g53b:cont_tau_n}. We overlaid these figures with curves representing constant time as prescribed by
\begin{equation}
    \tau = \sqrt{\frac{\mathrm{EM}_i}{1.2  V_i}} \times t,
\label{eq_g53b:time_curves}
\end{equation}
where $t$ represents the age of the remnant and EM$_i$ is emission measure $\int n_\mathrm{e} n_\mathrm{H} dV$. These figures further constrain the age of the plasma to $t_{\tau _0}\approx(5.2\pm1.0)d^{1/2}_{7.5}$~kyrs for region 1 and $t_{\tau _0}\approx(5.1\pm1.9)d^{1/2}_{7.5}$~kyrs for region 2 (in $1\sigma$ range). 

All of the estimation techniques used here have some caveats, which makes pinpointing the exact age of SNRs difficult.
However, combining all calculated estimates from regions 1 and 2
portrays a mostly consistent picture of the material being shocked around $\sim 1000 - 5000$~yrs ago. This estimate puts G53.41+0.03 itself into similar age range, and in a category of younger SNRs. 
The calculated extinction using \cite[ (Fig.~3)]{Predehl1995} points to high A$_v \sim 15$~mag, which would make it invisible to the observers of that time.

The range of possible ages calculated above deviates somewhat from the estimates reported in \cite{Driessen2018}, which were based on a larger estimate of the
radius. Moreover, \cite{Driessen2018} did not correct for the fact that the ionisation age, $t_{\rm \tau} \approx 0.3 t_{\tau _0}$ for SNRs expanding in a uniform medium, 
with $t_{\tau _0}$ giving the actual age of the SNR. 

\subsection{Point sources in the FoV}
\label{sec_g53b:discussion_point_sources}
The proximity to the Galactic plane and the location within the Sagitarius-Carina arm suggest that the SNR has a core-collapse origin. But a final confirmation needs a proper identification of one of the point sources as the stellar remnant, or more accurate abundance measurements associated with an ejecta component. 
The FoV around the SNR provides a number of point sources. Many of them are likely associated with the IRDC G53.2 and lie in the foreground of the SNR at $\sim 1.7$~kpc distance. \cite{Kim2015a} has identified $\sim$370 point sources from which about 300 of them are Young Stellar Object (YSO) candidates. We concentrate mainly on the study of the SNR, and we therefore do not go into detailed investigations of all of them. However, the point sources in the assumed geometrical centre of the SNR provide an intriguing hypothesis that they could be related to the SNR as its leftover co-created neutron star. We therefore examine them in more detail.

As the first step, we compared the positions of two point source locations (PS1: RA~19:29:56.0, DEC~+18:10:19.2; PS2: RA~19:29:54.3, DEC~+18:10:23.5) with the list of known YSOs from \cite{Kim2015a} and other potential SIMBAD registered sources. While the PS1 location (the lower of the two in Fig.~\ref{fig_g53b:SNR_regions}) did not provide any catalogued object within 24'', PS2 point source lies within 3'' of YSO identified as 2MASS J19295440+1810260. As this is within the resolving power of {\it XMM-Newton} (spatial resolution of $6''$ at 1keV) and YSOs are known X-ray emitters \citep{Giardino2007,Winston2010} they are most likely the same object and not associated with the SNR.

\begin{figure}
    \centering
    \includegraphics[width=\columnwidth]{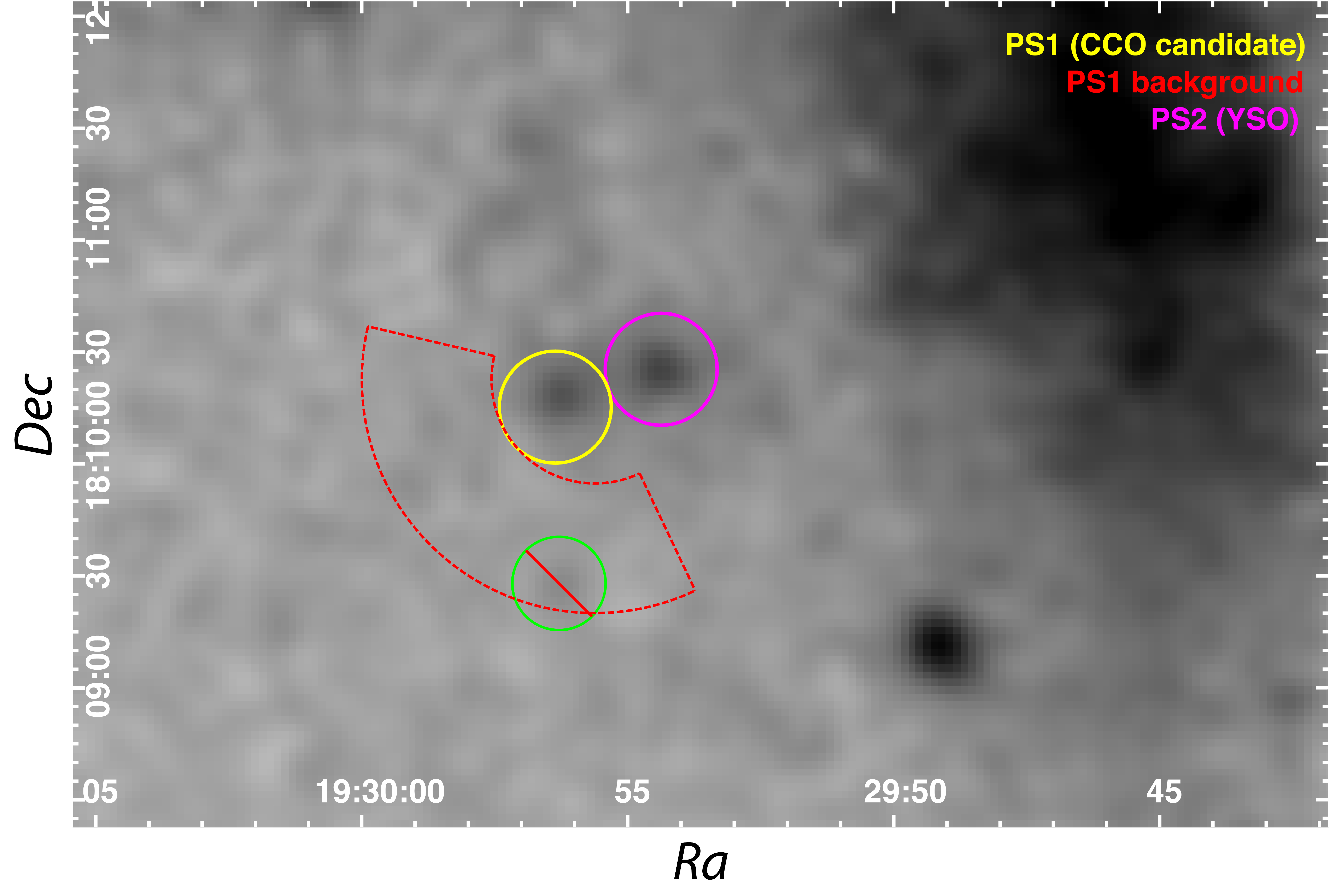}
    \caption{PS1 source and background extraction region.}
    \label{fig_g53b:PS1}
\end{figure}

To further investigate whether the PS1 central source could be a neutron star, we proceed with the extraction of the point source's spectrum and a nearby background in the 0.4--10.~keV energy range. As PS1 lies on different pn chip, which is unaffected by the pile up caused by the X-ray binary, we include the pn data in the point source analysis. We model the local background of all three detectors (Cstat/d.o.f of 186/182) and apply it to the binned (10 cts/bin) source spectrum. Thanks to the proximity of the background region to the source, we are able to freeze the Gaussian normalisation representing the instrumental lines as well. The extraction regions for the point source analysis are shown in Fig.~\ref{fig_g53b:PS1}.

\begin{figure*}
    \centering
    \includegraphics[width=\columnwidth]{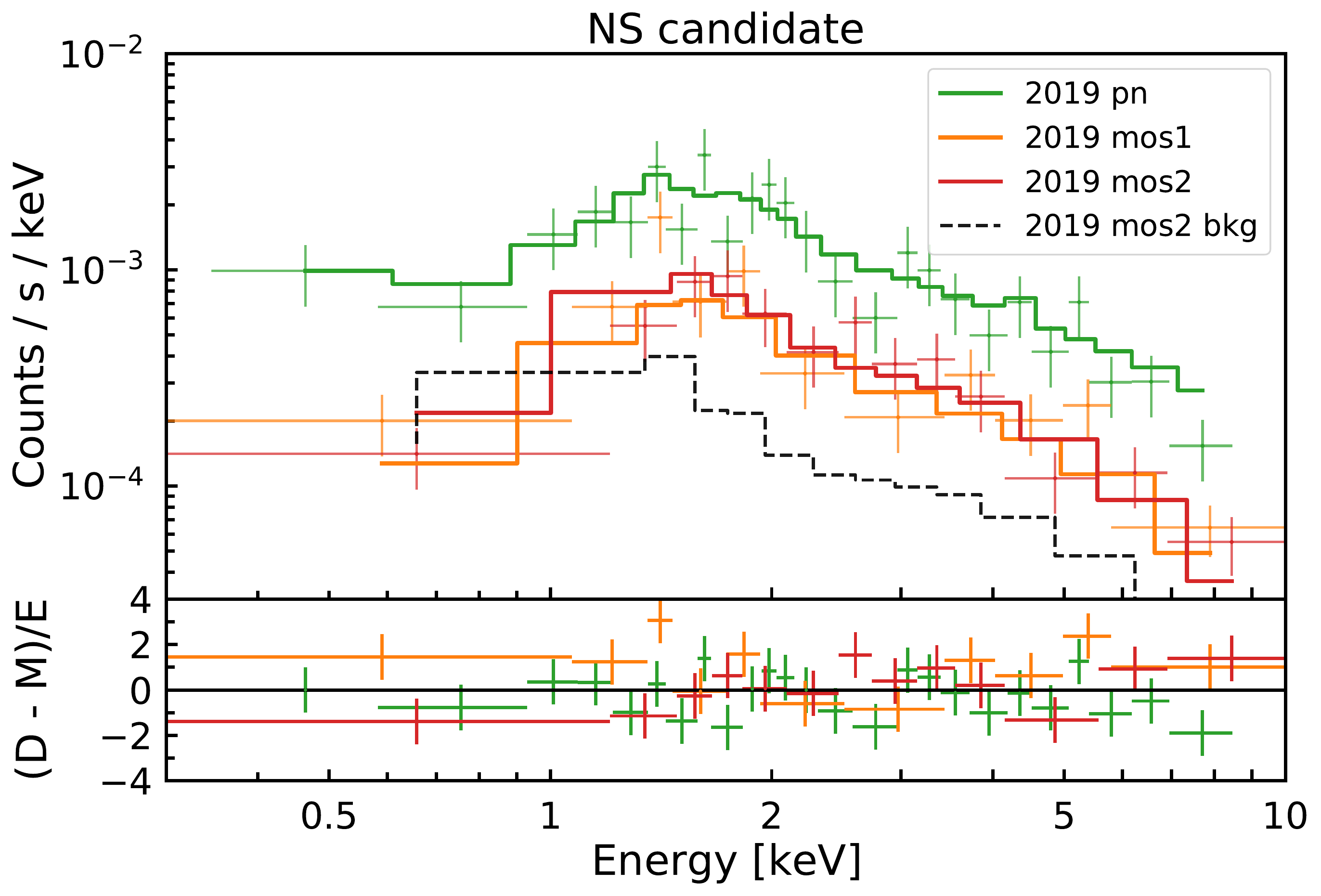}
    \includegraphics[width=\columnwidth]{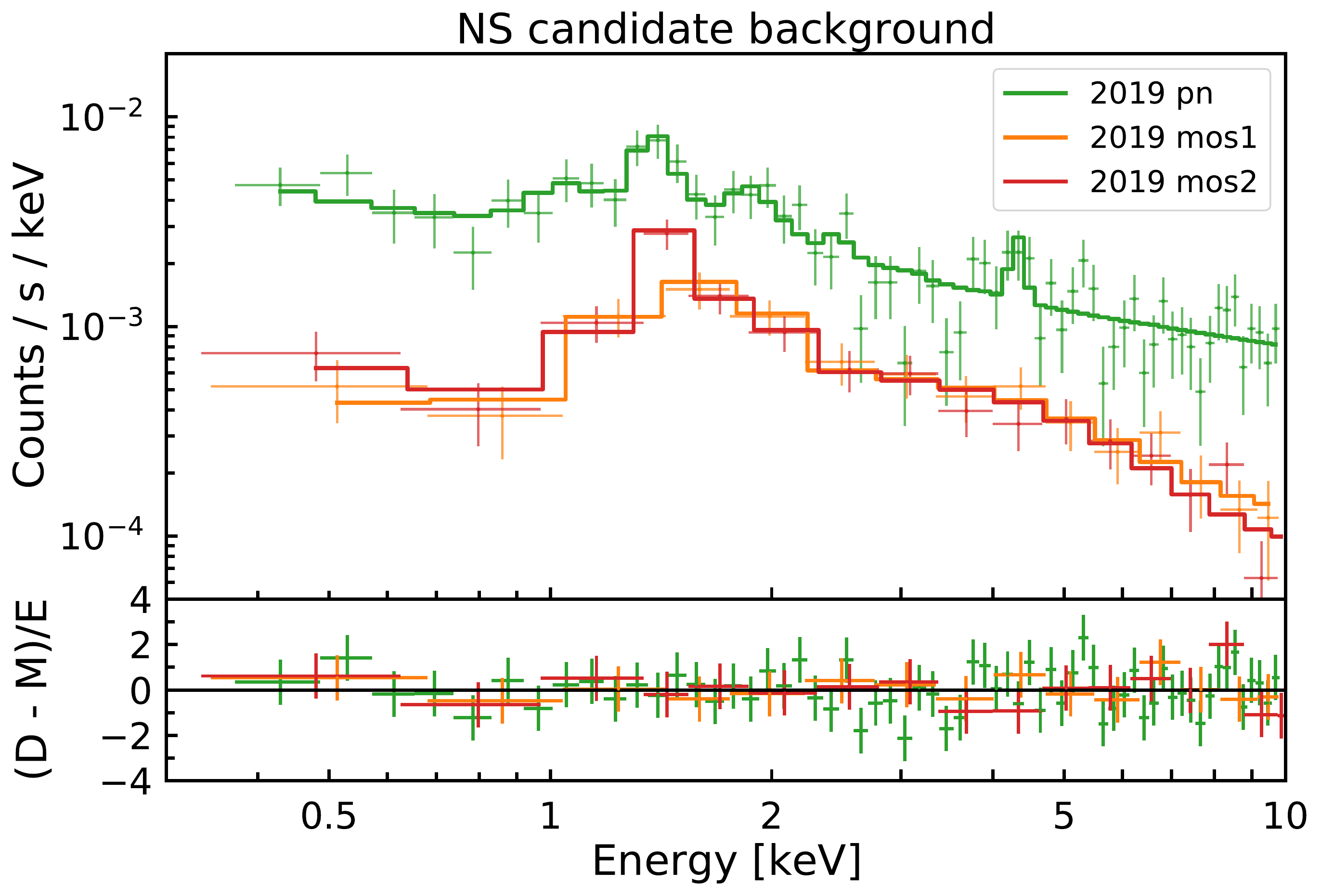}
    \caption{Source (left) and background (right) spectra for the PS1 neutron star candidate. The \textit{TBabs*pow} model is being displayed on top of the source spectra.}
    \label{fig_g53b:coo_candidate}
\end{figure*}

The spectrum of a young neutron star with an age similar to that of the SNR of SN $\sim 1-5$~kyrs ago could exhibit two radiation components: i) a thermal black-body component coming from the cooling surface of the neutron star and ii) a non-thermal power-law component produced for example by activity within the magnetosphere of the neutron star \citep{Potekhin2020}, or potentially a unresolved pulsar wind nebula. The properties of the neutron star depends on what category the neutron star belongs to: is it a young spin-down pulsar with surface magnetic field strength of $B\sim 10^{12}$~G, a low-magnetic field neutron star or central compact object (CCO) akin to the neutron star in Puppis A \citep[e.g.][]{Petre1996,Gotthelf2013}, or is it a high magnetic field ($B\sim 10^{14}-10^{15}$~G) magnetar, whose emission is powered by magnetic-field decay \citep[][for an overview]{Olausen2013}? CCO do not exhibit non-thermal emission, whereas typical spindown-powered pulsars and magnetars do have non-thermal X-ray radiation components.

For our neutron star candidate the spectrum shows a statistical preference for the non-thermal \textit{TBabs*pow} (55.5/46 Cstat/d.o.f.) over the thermal \textit{TBabs*bbodyrad} (60.6/46 Cstat/d.o.f.) model. Using both components at the same time (\textit{TBabs*(Bbodyrad*pow)} does not lead to a statistical improvement of the fit (55.4/44 Cstat/d.o.f.) and the parameters of \textit{Bbodyrad} remain unconstrained. Better quality of data could nevertheless still confirm it in the future. We present the tabulated results for the constrained models in table~\ref{tab_g53b:cco_candidate}.

The \textit{TBabs*Bbodyrad} model fits a high $\sim 1$~keV temperature compared to more common values of kT$<0.3$~keV reported for neutron stars \citep{Potekhin2020}. Furthermore, the fitted normalisation points towards an unreasonable emitting radius of R$\sim 3$~m. So most likely the point source is not a CCO.
On the other hand the estimated power-law index $\Gamma \sim 2$ in the \textit{TBabs*pow} model is consistent with typical values expected from the neutron stars, in particular young pulsars and magnetars \citep{Kaspi2006book, Olausen2013}. 
Note that the best-fit value of $N_{\rm H}$ is within 1$\sigma$ consistent with that of the SNR.
    
\begin{table}
    \centering
    \begin{tabular}{ll|cc}
    \hline
    Parameter & \multicolumn{1}{l|}{Units} & \textit{TBabs*bbodyrad} & \textit{TBabs*pow} \\
    \hline
    N$_\mathrm{H}$ & $\times 10^{22}$~cm$^{-2}$& 0.53$^{+0.5} _{-0.1}$  &1.8$^{+1.0} _{-0.1}$   \\
    kT & keV & 0.94$^{+0.06} _{-0.14}$ & - \\
    norm$_\mathrm{(bb)}$ & m & 2.88$^{+2.6} _{-0.6}$  & -  \\
    $\Gamma$ & & - & 2$^{+0.7} _{-0.1}$   \\
    norm$_\mathrm{(pow)}$ & $\times 10^{-5}$ & - & 1.1$^{+1.5} _{-0.1}$ \\
    \hline
    \multicolumn{2}{l|}{Cstat/(d.o.f.)} & 60.6/46 & 55.5/46  \\
    \end{tabular}
    \caption{Best fit results of the PS1 point source in geometrical centre. 1$\sigma$ range was calculated using XSPEC's MCMC method with 10$^6$ steps and 600 walkers.}
    \label{tab_g53b:cco_candidate}
\end{table}

The point source is also faint (F$_\mathrm{pow, 1-10~keV} = 3.7\times 10^{-14}$~erg~cm$^{-2}$s$^{-1}$) and at distance of 7.5~kpc has a luminosity of $L_\mathrm{1-10~keV} \approx 2\times 10^{32}$~erg~s$^{-1}$. If we take a young pulsar as a potential candidate for this source and place it in the $P\Dot{P}$ diagram \citep[ Fig~6.3]{Condon2016} it would occupy space of known magnetars. This indicates that the point source is not luminous enough to be a young spindown powered pulsar.
However, it would still allow for the possibility that the point source is a magnetar, which typically also exhibit a power-law radiation component in X-rays, but with radiation powered by magnetic activity. 

The lack of strong evidence for a thermal component and inability to do a temporal analysis due to low count rate does not allow us to definitively conclude on the neutron star/magnetar nature of this point source. But even short exposure investigations with future observatories like Athena could potentially answer this question.
Moreover, if the point source is indeed a magnetar there is the possibility that it will at some point exhibit magnetar flares.

During the 2019 observation we detected a type I outburst of a high mass x-ray binary IGR J19294+1816 \citep{Domcek2019ATel}. The spectrum shows a similar absorption as region 1 and 2 of the SNR. \cite{Rodriguez2009} has placed this binary to a distance of about 8~kpc. While it is not directly associated with the SNR, as both objects are in the end stage of their stellar evolution, they could potentially belong to the same star association.

Furthermore, we found two additional X-ray transients unrelated to the SNR (Ra~19:29:37.6, Dec~+18:08:49.8 and Ra~19:29:29.3, Dec~+18:05:12.4). The light curves and tentative spectral analysis point towards M-dwarfs experiencing flaring events. However, more detailed analysis is beyond the scope of this paper. 

\section{Conclusions}
We investigated a recently discovered SNR G53.41+0.03 with a new dedicated {\it XMM-Newton} observation. Our main findings are:
\begin{itemize}
    \item The X-ray morphology of the SNR is consistent with the detected radio structure in \cite{Driessen2018} and consists of an incomplete shell with the southern part being either below the detection threshold or missing. 
    \item There are two unique regions of the remnant with significantly detected emission, characterised by the non-equilibrium ionisation plasma model. The difference in their brightness and plasma characteristics is likely caused by the higher density in the brighter region and a combination of lower density and the proximity to Galactic plane in the fainter region.
    \item The spectral analysis reveals the age of the SNR to be likely in range of $1000-5000$~yrs.
    \item We find two point sources in the geometrical centre of the remnant, one of which we confirm as a young stellar object (YSO). The other point source has characteristics consistent with a magnetar, but further investigations are required to confirm its nature.
\end{itemize}

\section{Data availability}
The data underlying this article are available in the zenodo repository, at \url{https://doi.org/10.5281/zenodo.4737383}.

\begin{acknowledgements}
    We would like to thank N.~D.~Degenaar, R.~A.~D.~Wijnands for discussions on the topic of neutron stars, G.~Mastroserio, R.~M.~Connors, J.~V.~Hernández~Santisteban for advise on XSPEC's MCMC and XMM helpdesk team for their technical support. We would also like to thank the anonymous referee for their comments that greatly improved this paper.
    The work of VD is supported by a grant from NWO graduate program/GRAPPA-PhD program. PZ acknowledges the support from the NWO Veni Fellowship, grant no. 639.041.647 and NSFC grant 11590781. LND acknowledges support from the European Research Council (ERC) under the European Union's Horizon 2020 research and innovation programme (grant agreement No 694745).\\
    
    This research made use of {\sc astropy}, a community-developed core {\sc python} package for Astronomy \citep{Robitaille2013}, {\sc matplotlib} \citep{Hunter2007} and {\sc corner} \citep{Foreman-Mackey2016}. We further made use of {\sc SAOImage DS9} \citep{Joye2003}, {\sc XSPEC} \citep{Arnaud1996} and the SAO/NASA Astrophysics Data System. 
\end{acknowledgements}

%
%
\bibliographystyle{aa}
\bibliography{g53b}

\appendix
\section{MCMC results for SNR regions 1, 2 and 3}
\begin{figure*}
    \centering
    \includegraphics[width=\linewidth]{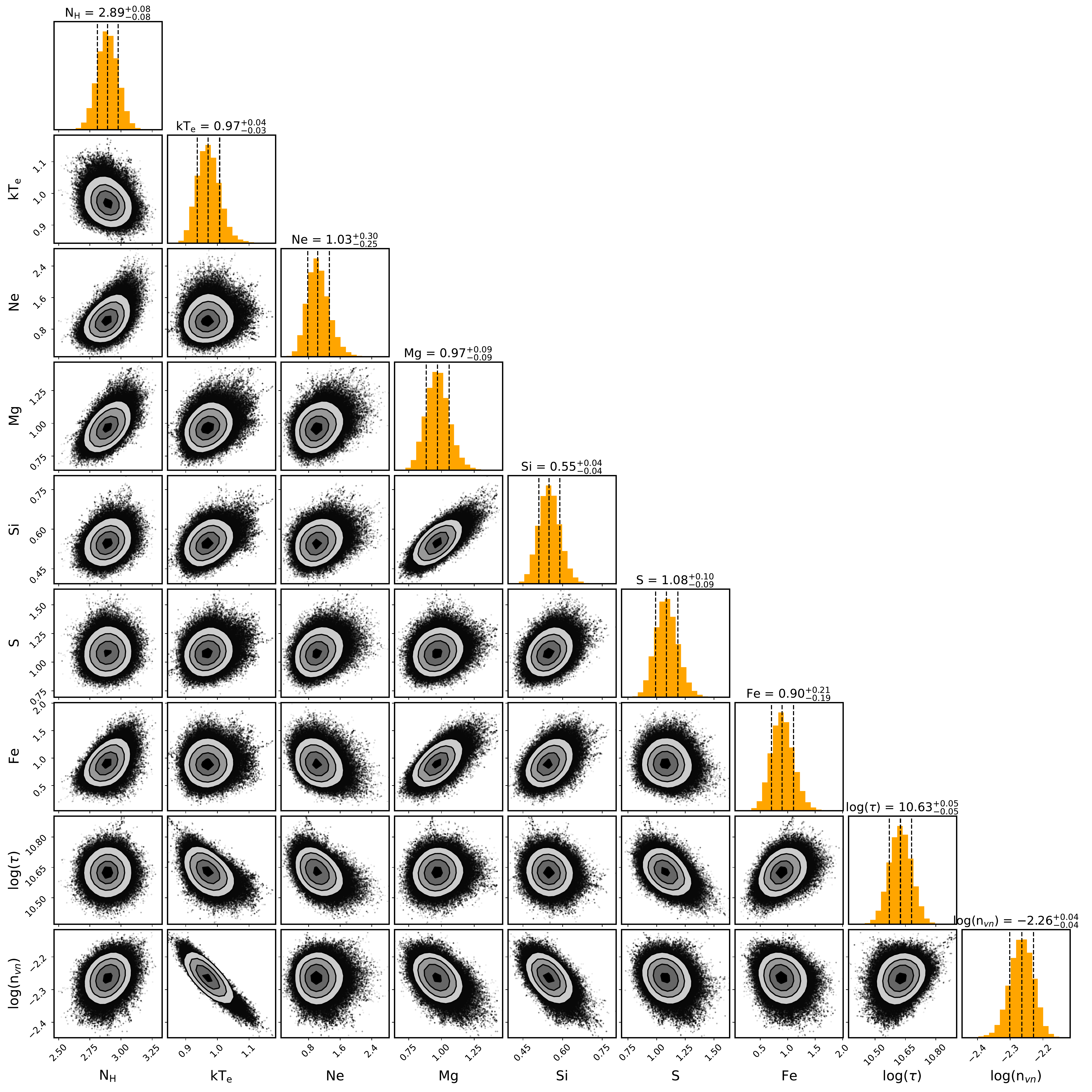}
    \caption{Two-dimensional correlation distributions of the \textit{TBabs*vnei} part of the model for region 1 of the SNR G53.41+0.03. Posterior distributions of individual parameters are described by the median and the 1$\sigma$ confidence interval. Contours in the 2D histogram represent 0.5, 1, 1.5 and 2$\sigma$ levels. Instrumental line components are not displayed for readability reasons. Full version of the figure, including the instrumental line components, is available in the \href{https://doi.org/10.5281/zenodo.4737383}{zenodo} archive.}
    \label{fig_g53b:mcmc_region1}
\end{figure*}

\begin{figure*}
    \centering
    \includegraphics[width=\linewidth]{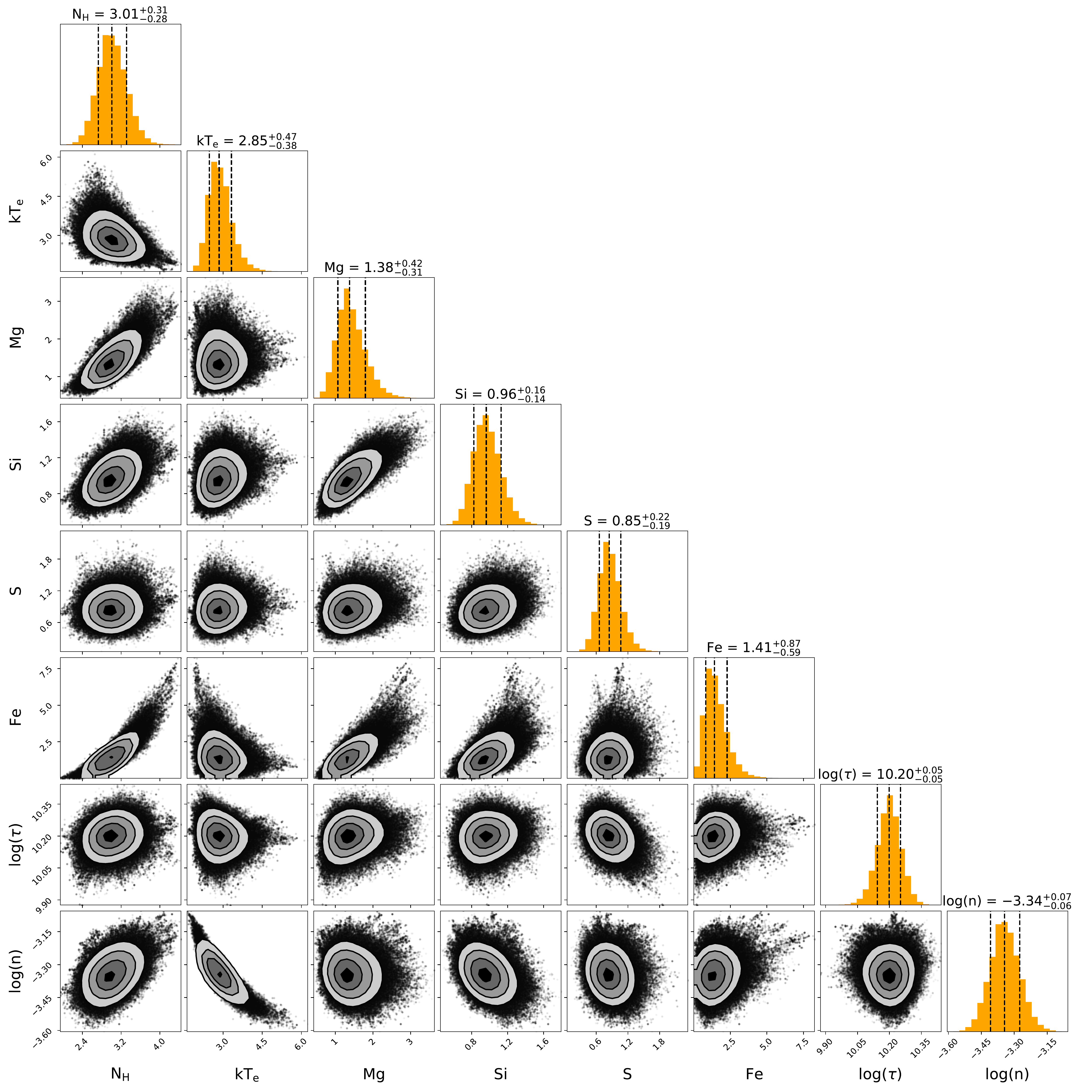}
    \caption{Same as in Fig.~\ref{fig_g53b:mcmc_region1} but for model \textit{TBabs*vnei} region 2.}
    \label{fig_g53b:mcmc_region2}
\end{figure*}

\begin{figure*}
    \centering
    \includegraphics[width=\linewidth]{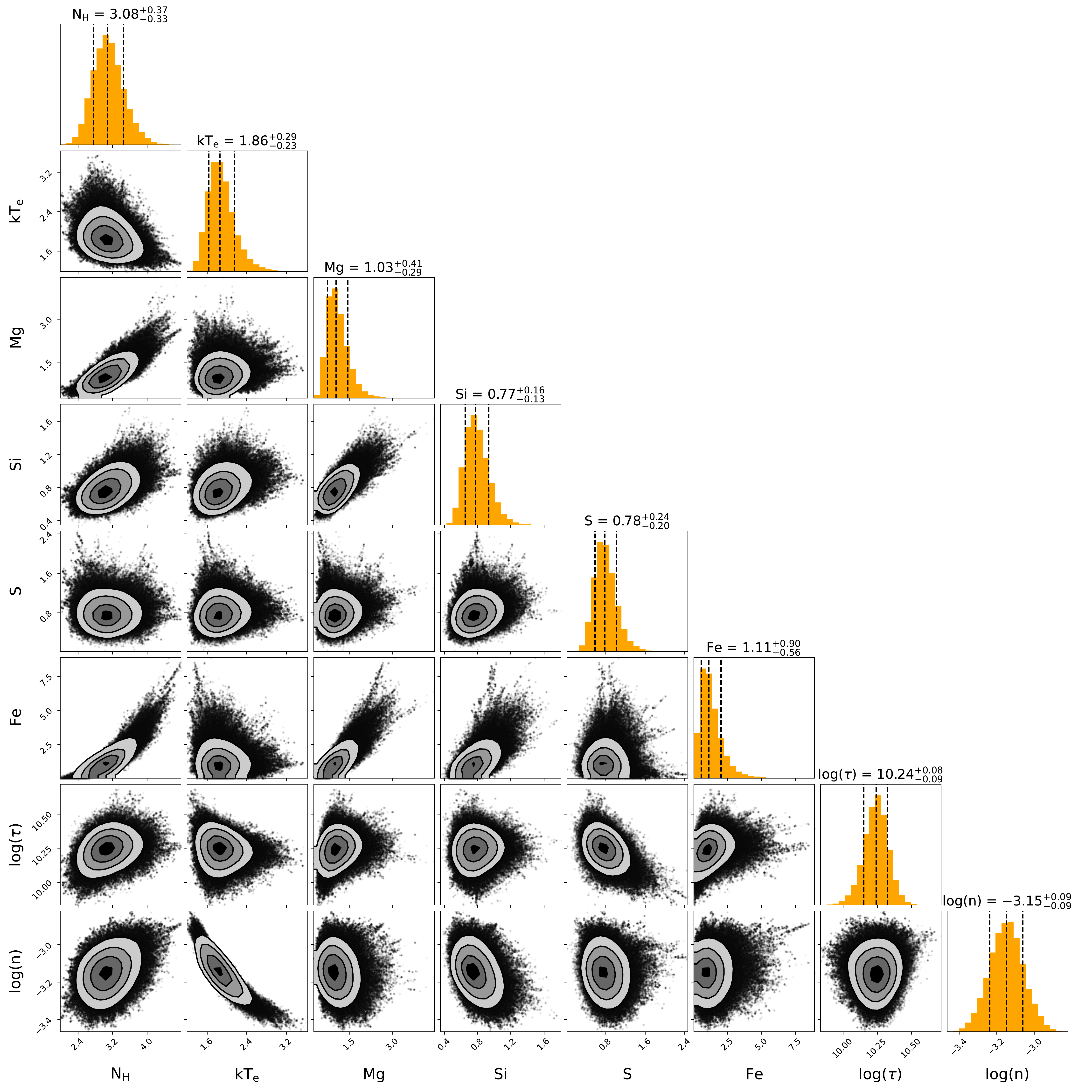}
    \caption{Same as in Fig.~\ref{fig_g53b:mcmc_region1} but for model \textit{TBabs*vnei} (5keV cut) of region 2.}
    \label{fig_g53b:mcmc_region2_2}
\end{figure*}

\section{MCMC results for the neutron star candidate PS1}

\begin{figure*}
    \centering
    \includegraphics[width=\columnwidth]{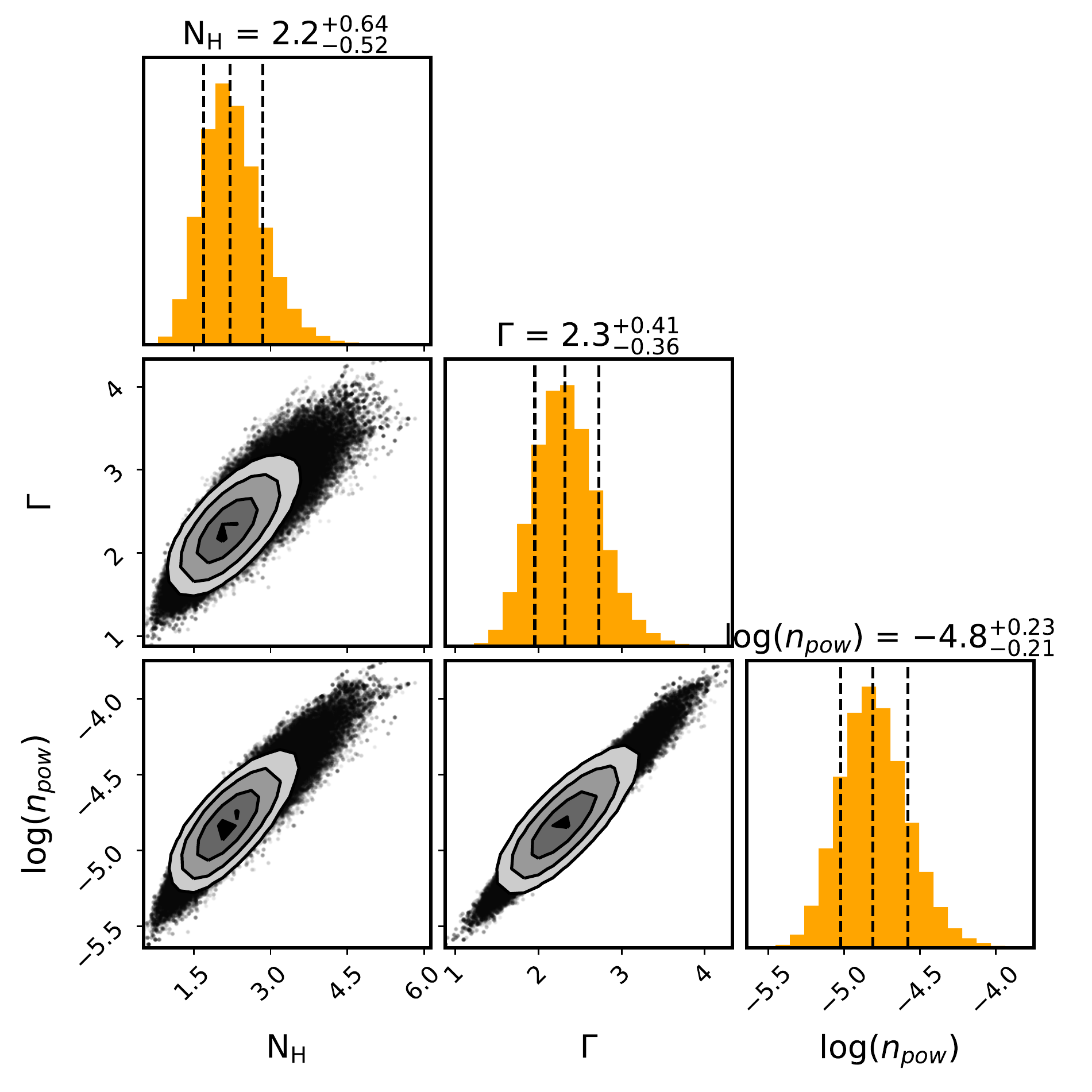}
    \includegraphics[width=\columnwidth]{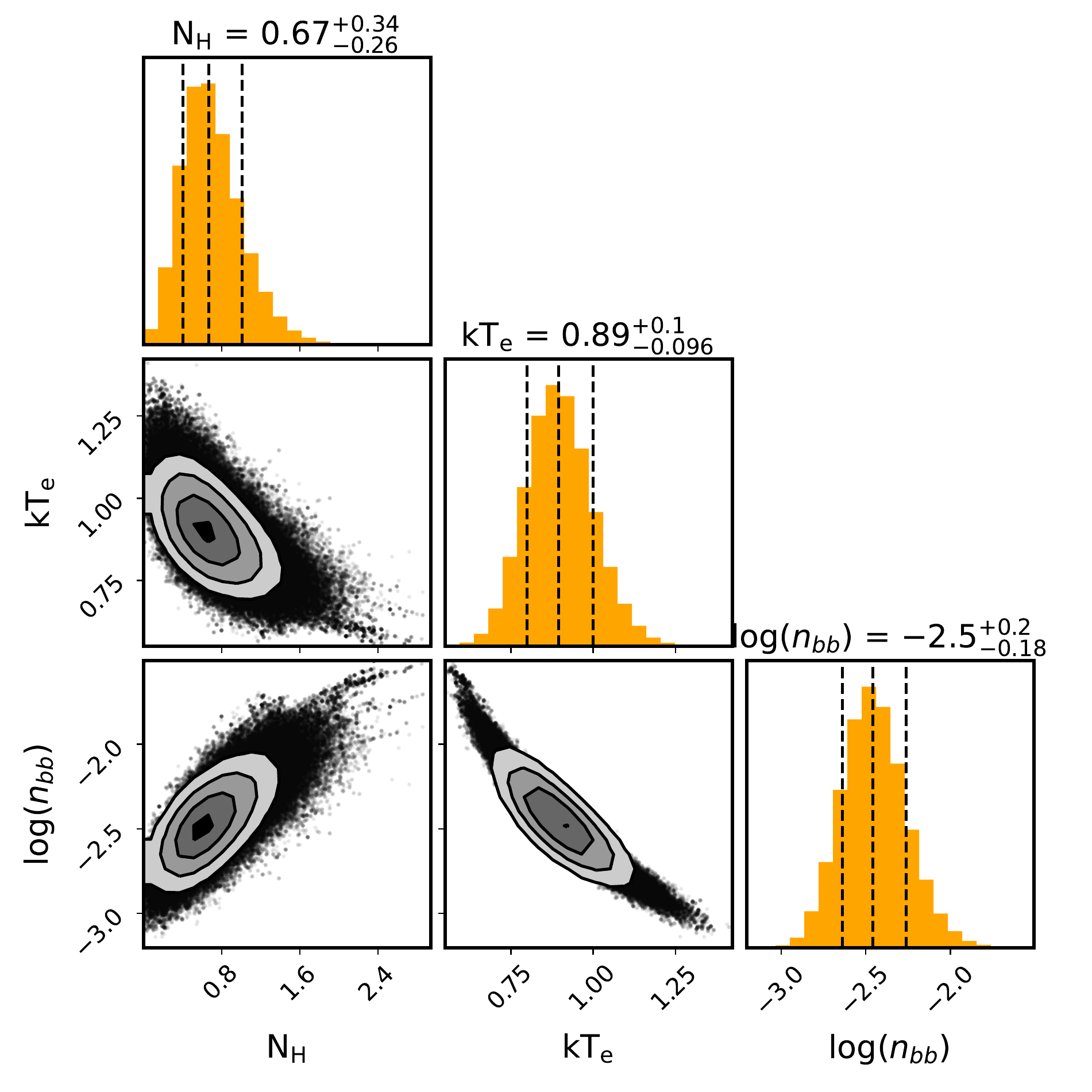}
    \caption{Two-dimensional correlation distributions of the \textit{TBabs*pow} (left) and \textit{TBabs*bbodyrad} (right) models for the PS1 neutron star candidate. Posterior distributions of individual parameters are described by the median and the 1$\sigma$ confidence interval. Contours in the 2D histogram represent 0.5, 1, 1.5 and 2$\sigma$ levels. }
    \label{fig_g53b:mcmc_ps1}
\end{figure*}

\end{document}